# Influence of the shape of RDX grains on the creation of hot spots by mesoscale modeling

*Steve Belon[1], Benjamin Erzar[1], Elodie Kaeshammer[1], Lionel Borne[2]*

steve.belon@cea.fr

[1] *CEA, DAM, GRAMAT, P.O. Box 80200, 46500 Gramat, France.*

[2] *ISL (French-German Research Institute of Saint Louis), 5 rue du Général Cassagnou, P.O. Box 70034, 68301 Saint-Louis Cedex, France*

Abstract

CEA-Gramat studies the sensitivity of energetic materials to enhance their security and reliability. The conditions leading to the initiation of an explosive must be understood to control its sensitivity. According to the hot spots theory, the shock initiation of heterogeneous explosives is related to their microstructure: the shock interacts with the heterogeneities of the microstructure (pores and inclusions, morphology of grains and fragments, debonding, etc.) and creates local deposits of energy. To describe these hot spots, energetic materials have to be modeled at a scale allowing the discretization of their microstructure: the mesoscale.

Micro-computed tomographies of energetic materials are done at CEA-Gramat and analyzed to build geometric models used in finite element simulations. Two kinds of models are studied:
- Real models are directly built on the real microstructures extracted from micro-computed tomographies.
- Virtual models are based on the same microstructures but simplified to study independently the effects of microstructural parameters (granulometry, porosity, filler content…) on the creation of hot spots.

Compositions based on different kind of RDX particles in an inert binder are studied through numerical simulation. The influence of particle shape on the inert shock response is investigated at the mesoscale. Local heterogeneities of pressure and temperature fields appear intimately related to the morphological properties of the microstructures. Particles with sharp edges create more hot spots than spherical particles.

Introduction

The shock-initiation of heterogeneous solid explosives is related to the formation of hot spots. Hot spots depict localized elevation of pressure and temperature in the material. They are due to the interaction of the propagating shock with the heterogeneities of the material microstructure. According to the size, intensity and density of hot spots, they will either extinguish or be the starting point of chemical reactions. Knowing the formation of hot spots is thus of primary importance to handle the initiation of solid explosives.

Numerous studies concerning the formation of hot spots have been published; see for instance [1, 2]. Several mechanisms have been identified to be a potential source of hot spots, including void collapse, plastic deformation, fracture of grains, friction, debonding [3, 4, 5, 6]. Each of these mechanisms may play a role in the appearance of hot spots, according to the kind of energetic material or the kind of loading considered. A better knowledge of the effects of each of these mechanisms allows a better control of the initiation of solid explosives, whether for intentional or accidental initiation. All these mechanisms have in common to be strongly dependent on the microstructure of the energetic material. Thus, the initiation of heterogeneous explosives should be studied at a scale taking into account their microstructural parameters.

The energetic material studied in this work is a plastic-bonded explosive (PBX) made of 70 % in weight of RDX and 30 % in weight of wax. The considered loading is an impact of a rigid projectile at 1200 m/s, corresponding to the initiation threshold of the RDX/Wax composition. The scenario studied in this work thus corresponds with an accidental kind of initiation. The microstructural parameter examined in this paper is the effect of the shape of RDX grains on the appearance of hot spots. Influence of the shape of RDX crystals on the shock-sensitivity of PBXs has been observed experimentally; see for instance van der Steen *et al.* [7] who noticed a greater sensitivity threshold of spherical particles than irregularly shaped particles. The objective of the present work is to show the influence of the RDX crystal shape on the propagating shock locally, thanks to mesoscale numerical modeling.

Materials

The energetic materials studied in this work are RDX/Wax compositions produced by the french-german research Institute of Saint-Louis (ISL). They are composed of 70 % in weight of RDX particles and 30 % in weight of wax. The RDX grains used in these compositions are subjected to two treatments to become less sensitive than common grade RDX particles. This process, presented at Figure 1, is a trademark of ISL [8]. The first treatment consists in a recrystallization of commercial grade RS-RDX particles furnished by Eurenco. Most of crystal internal defects, including porosities and micro-cracks, disappear during the recrystallization. The resulting particles have less than 0.05 % in volume of internal voids. The shape of the particles is also modified by the recrystallization process as they present sharp edges, whereas the RS-RDX particles are rather spherical. Recrystallized particles are called "raw VI-RDX". The second step is a surface treatment process aiming at removing the sharp edges created by the recrystallization. The resulting particles are spherical and called "VI-RDX".

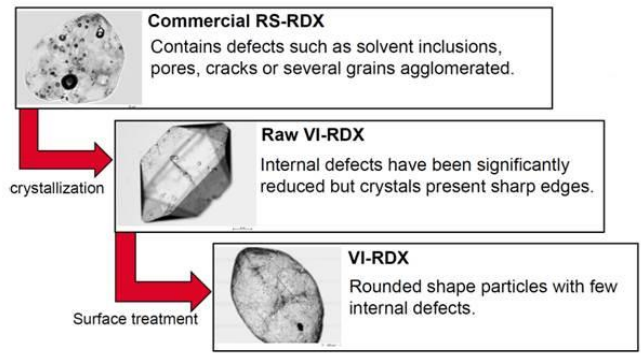

Figure 1: Steps of the formation of VI-RDX crystals

ISL produced RDX/Wax compositions using either raw VI-RDX or VI-RDX particles. They characterized the shock-sensitivity threshold of these compositions by impact experiments [9]. Their results are presented at Figure 2. They found that the sensitivity threshold of the raw VI-RDX composition (6.2 GPa) is lower than the pressure threshold of the VI-RDX composition (7.2 GPa). The only difference between both compositions is the shape of the RDX particles. All other microstructural parameters are identical: weight and size distributions of RDX grains, rate of porosity, etc. Thus, the difference in the sensitivity threshold of both compositions under shock-loading should be explained by the effect of the RDX crystal shape on the propagating shock.

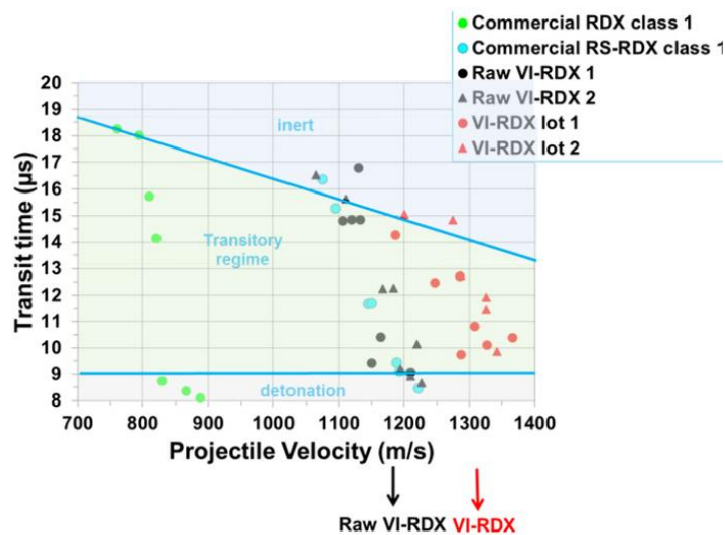

Figure 2: Results of ISL small projectile impact experiments for several formulations

Methods

To understand the effect of the shape of the raw VI-RDX and VI-RDX grains on the propagating shock we model the microstructure of both compositions at the mesoscale. First, we model RDX crystals with 2D geometrical forms: the angular raw VI-RDX particle is represented by a square and the spherical VI-RDX grain is represented by a disc. These models contain only one grain in the domain to study the interaction of the propagating shock and the grain without disturbance coming from other grains. Secondly, we model both raw VI-RDX and VI-RDX compositions with 3D realistic microstructures to quantify the appearance of hot spots in presence of many reflective shocks.

Our models run with the eulerian OURANOS solver, developed at the CEA, using a BBC integration scheme. RDX is modeled with a Birch-Murnaghan equation of state and a Johnson-Cook behavior taking into account a linear thermal softening. RDX is treated as an inert: the chemical initiation is not modeled. The wax material is also modeled with a Birch-Murnaghan equation of state. Wax is considered purely hydrostatic as the melting temperature of this binder is low (350 K at ambient pressure). The shock loading represents a steel projectile continuously moving at the speed of 1200 m/s. This is modeled by an inflow boundary. Reflective boundary conditions, similar to rigid walls, are on the top and bottom sides of the domain. The 2D models span on 900x900 µm² with a constant mesh size of 1 µm, totalizing 810,000 elements. The raw VI-RDX particle is represented by a square with a side of 230 µm. The VI-RDX grain is represented by a disc with a diameter of 260 µm. The surfaces of the square and the disc are identical, so that both models are comparable. The simulation time of the 2D model is 200 ns, corresponding to the time necessary for the shock to propagate along the whole domain. Materials and boundary conditions of the 3D models are identical to those of the 2D models. The 3D models span on 1x1x2.5 mm³ with a constant mesh size of 3.33 µm, totalizing 67.5 million cells. The simulation time of the 3D model is 600 ns.

To compare the shock-sensitivity of the different configurations we have to quantify the appearance of hot spots. There is no standard definition of a hot spot. Several criteria have been suggested to characterize the existence of hot spots; see for instance [10, 11, 12]. They are commonly based on a pressure, temperature or energetic threshold. We choose a hot spot criterion resting on a temperature threshold: the melting temperature of RDX crystal. At ambient pressure RDX grains melt at 500 K. We assume an evolution of the melting temperature with density following a Lindemann type law. The loading considered in the present work leads to a melting temperature around 600 K for RDX crystals.

Results of the 2D models

In 2D, we model RDX grains with geometrical forms to simplify the observation of the interaction between the propagating shock and the grain. Angular raw VI-RDX particles are represented by a square and spherical VI-RDX particles are modeled with a disc. The time sequences of temperature of these configurations are presented at Figure 3. The orientation of the square to the propagating shock varies from 0 to 45 degrees to observe if any difference occurs in the interaction of the shock with the grain.

We notice that the first interaction of the lead shock with the grain (t = 95 ns) is not sufficient to reach the hot spot criterion. As the shock passes, a Mach stem forms at the top and bottom sides of the disc and the square oriented at 22.5 and 45 degrees (t = 110 ns). These Mach stems are the places where hot spots appear. The Mach stems keep heating the same surfaces of the grains as their triple points move along the lead front (t = 125 ns). The square particle oriented at 0 degree does not produce Mach stems. In this configuration the temperature never reaches the hot spot threshold. In the absence of microstructural defects inside the grains (porosities, cracks, etc.) no hot spot appears inside the particle (t = 140 ns). These results are similar to those obtained by Menikoff [13] who studied numerically the formation of hot spots by shock reflections in nitromethane charged with glass beads.

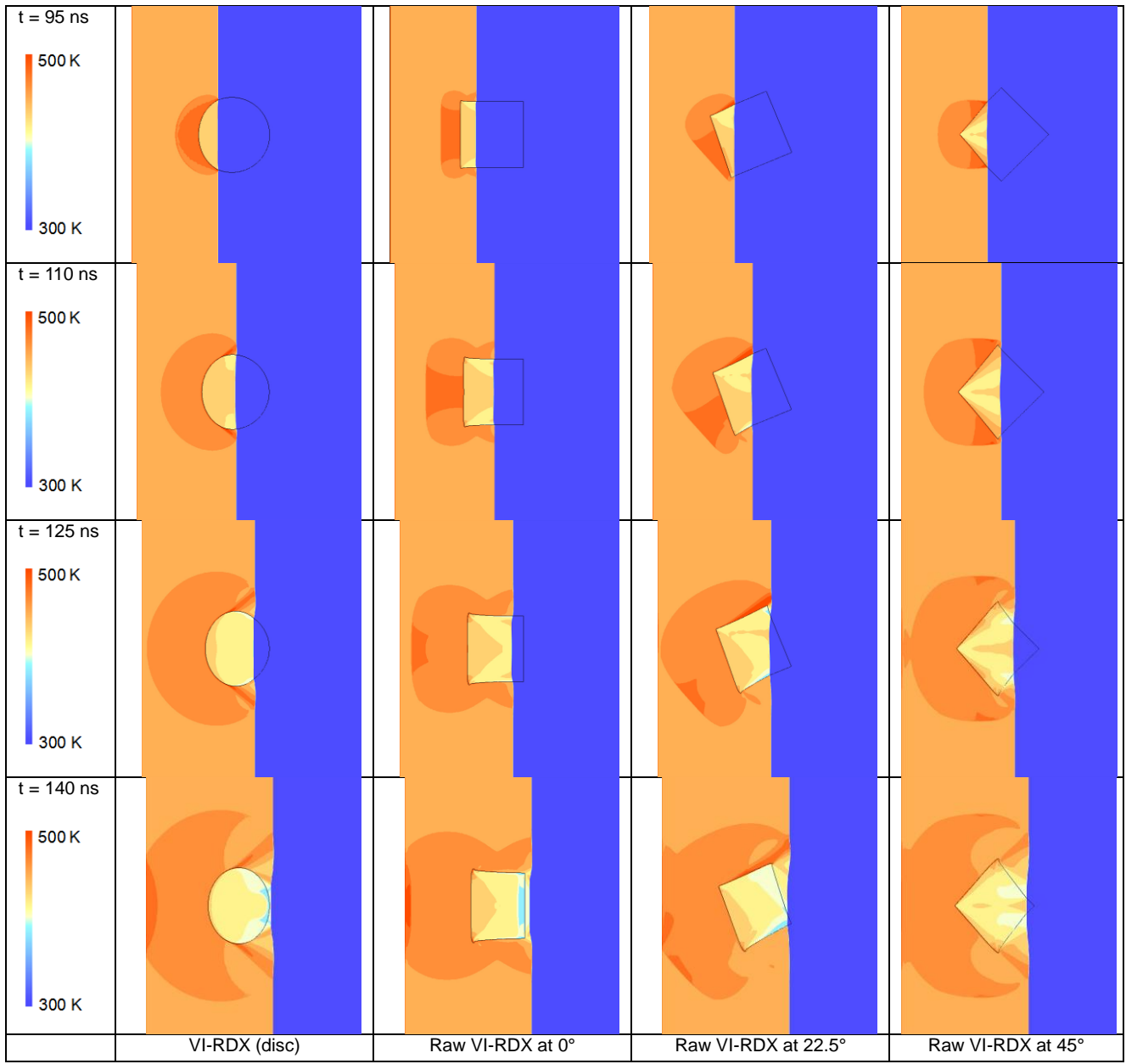

*Figure 3: Time sequences of temperature for the VI-RDX particle (disc) and the raw VI-RDX particle (square oriented at 0, 22.5 and 45 degree)*

Results of the 3D models

In 3D we are interested in studying the effect of shock reflections inside a real microstructure on the formation of hot spots. Thus, we model real grains of raw VI-RDX and VI-RDX in a large domain respecting the 70/30 weight ratio between charges and binder. The intra-granular and extra-granular porosities of these compositions are especially low. Yet, no porosity is included to be sure that the only parameter influencing the shock propagation is the RDX crystal shape. Besides removing any porosity, the microstructures are simplified by using only one grain of each kind. In other words, we select one real grain of raw VI-RDX, representative of the mean microstructural properties of this kind of grain, and build a composition made of a hundred copies of this grain. The same process is done with a mean VI-RDX particle to obtain a composition made of a hundred copies of the particle. This simplification of the microstructure avoids the effects of the variation of particle size and shape inside one kind of composition, which is not the goal of the present work. Figure 4 and Figure 5 depict the selected real grains, their finite element models and the representative microstructures made of these particles.

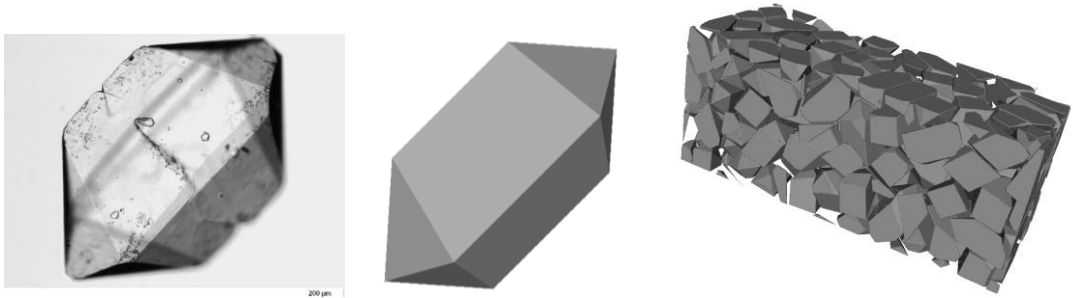

Figure 4: The raw VI-RDX crystal modeled to build the microstructure

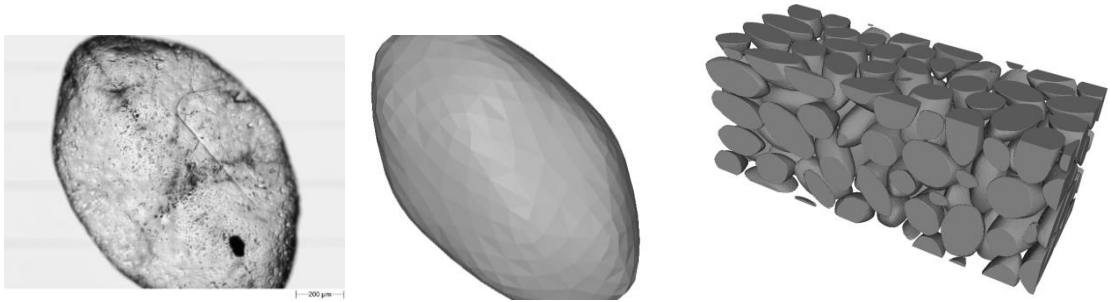

Figure 5: The VI-RDX crystal modeled to build the microstructure

Figure 7 presents the time sequence of temperature of both raw VI-RDX and VI-RDX compositions. It is difficult to observe locally the influence of particle shape on the propagating shock in a real microstructure because of the numerous shock-particle interactions and shock-wave reflections. Yet, we notice that the angular raw VI-RDX particles cause more hot spots than the spherical VI-RDX particles. The hot spots formed in the raw VI-RDX material are also of greater intensity than in the VI-RDX material.

To compare results obtained on both compositions we focus on the hot spots extracted from these models. Figure 6 depicts the time evolution of hot spot density for both microstructures. In the raw VI-RDX material hot spots start to appear from the beginning of the simulation whereas for the VI-RDX material hot spots start to form at a third of the simulation. The rate of appearance of hot spots is slightly greater for the raw VI-RDX than the VI-RDX composition. At the end of the simulation there is 50 % more hot spots in the microstructure built with angular raw VI-RDX particles than in the microstructure built with spherical VI-RDX particles.

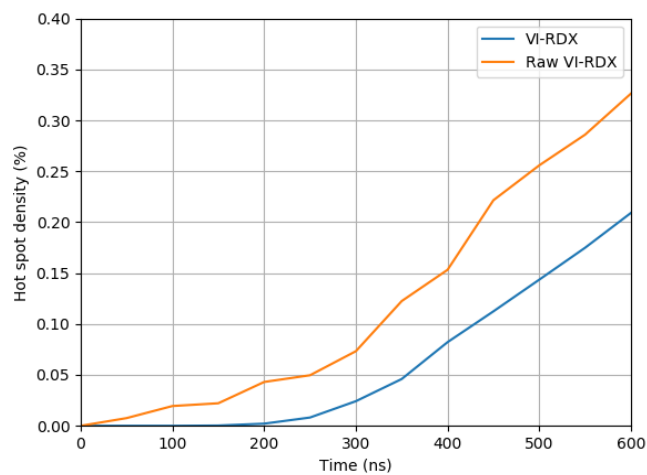

Figure 6: Hot spot densities of raw VI-RDX and VI-RDX compositions

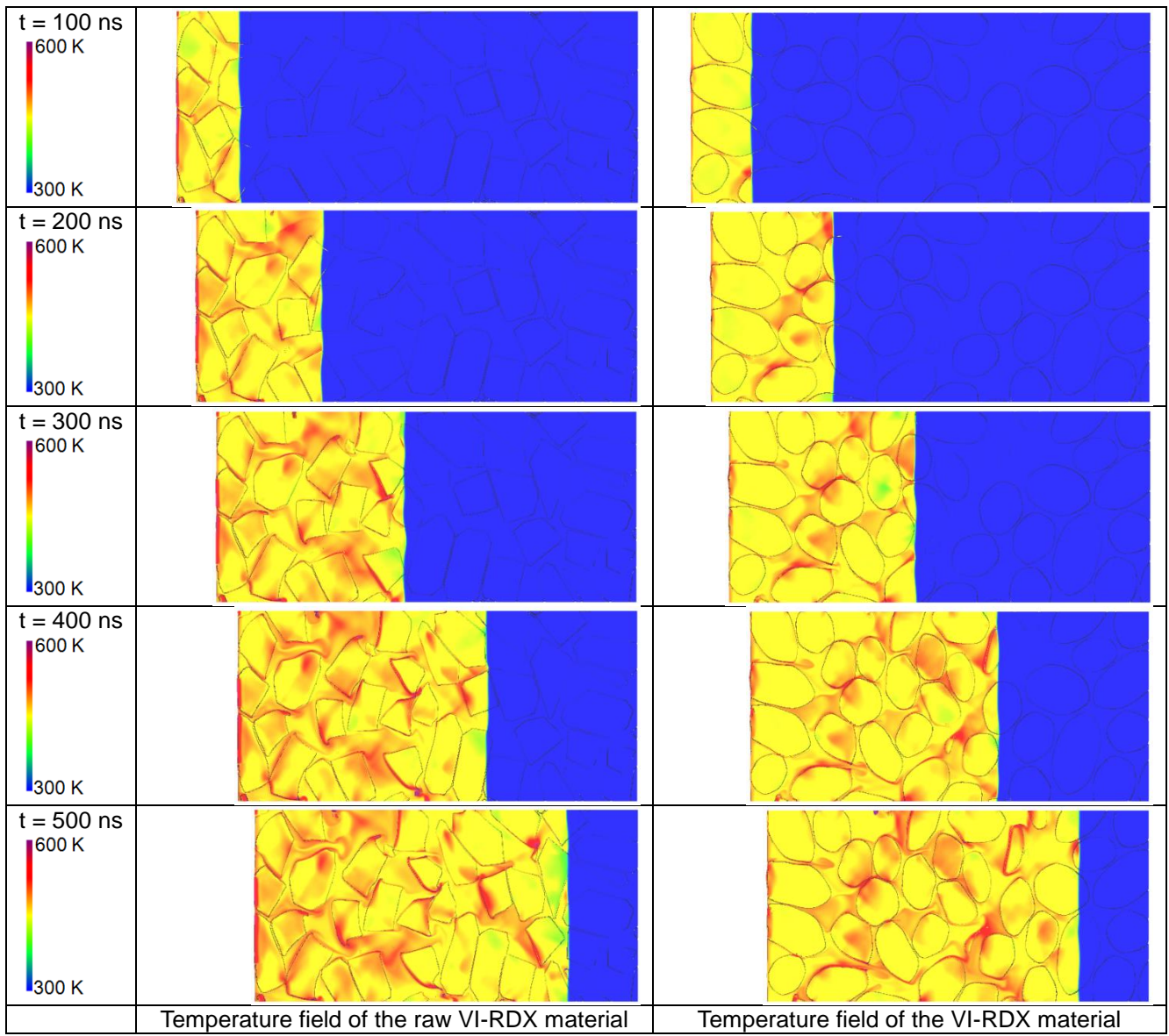

*Figure 7: Time sequence of temperature for both raw VI-RDX and VI-RDX compositions*

Discussion

The models used to perform the present comparison of the shock-sensitivity of angular and spherical RDX crystals suffer of several limitations. They are lacking of mechanisms like thermal conduction, grains fracture, plastic heating, etc. The quantification of hot spots is not accurate as we sum the cells reaching a temperature threshold taking into account neither the lighting duration nor the merging of hot spots. Moreover the microstructures used in this study are free of all other kinds of defects like porosities, which have been identified to be a source of hot spots.

Nevertheless this work highlights some mechanisms involved in the creation of hot spots. The formation of a Mach stem at the surface of the explosive particles seems of primary importance for the initiation of a single crystal as only the configurations where a Mach stem forms reach the hot spot criterion.

In the 3D realistic microstructures it is hard to say if hot spots appear from a Mach stem similarly to what occurs in 2D models. The difference in the amount of measured hot spots between angular and spherical crystals is sufficiently large to claim that the shape of explosive particles by itself is an important parameter for the shock-sensitivity of heterogeneous energetic materials.

## Conclusion

Influence of RDX crystal shape on the formation of hot spots in a shocked plastic-bonded explosive is studied by numerical modeling at the mesoscale. This work is based on two RDX/Wax compositions produced by ISL. One contains RDX crystals presenting sharp edges. The other contains only rounded particles. Two kinds of models are used to compare the effects of angular and spherical grains.
2D models made of only one grain in a wax binder are used to study the interaction between the propagating shock and the grain. Grains are assumed free of internal defects and modeled with simple geometrical forms. In these conditions hot spots only appear on the sides of the grain where a Mach stem forms. This location is related to the shape of the grain and to the orientation of the grain with the leading shock front.
3D models made with a hundred of grains are used to study the formation of hot spots in a real material microstructure subject to many shock reflections. Shapes of real explosive grains are used in the 3D models. Other microstructural parameters are not considered, especially porosity. The composition made of angular RDX crystals creates twice more hot spots than the material made of spherical RDX particles.

## Acknowledgements

This work is part of a collaboration between ISL and CEA-Gramat. It is supported by the French General Directorate for Armament (DGA) of the Ministry of the Armed Forces.